\documentclass[conference]{IEEEtran}
\IEEEoverridecommandlockouts
\usepackage{cite}
\usepackage[utf8]{inputenc} 
\usepackage[T1]{fontenc}
\usepackage{url}
\usepackage{ifthen}
\usepackage{amsmath,amssymb,amsfonts}
\usepackage{algorithmic}
\usepackage{algorithm}
\usepackage{subcaption}
\usepackage{mathtools}
\DeclarePairedDelimiter\abs{\lvert}{\rvert}
\usepackage{caption}
\usepackage{graphicx}
\usepackage{textcomp}
\usepackage{xcolor}
\newtheorem{remark}{Remark}
\usepackage[bookmarks=false]{hyperref}

\newcommand{\pnext}{.\textit{nx}}
\newcommand{\pprev}{.\textit{pr}}
\newcommand{\msf}{\mathsf}
\input{widebar_hack.tex}
\def\BibTeX{{\rm B\kern-.05em{\sc i\kern-.025em b}\kern-.08em
    T\kern-.1667em\lower.7ex\hbox{E}\kern-.125emX}}
\begin{document}
\begin{NoHyper}
\title{A Reinforcement Learning Approach for Scheduling in mmWave Networks}

\author{
\IEEEauthorblockN{Mine Gokce Dogan$^\dagger$, Yahya H. Ezzeldin$^*$, Christina Fragouli$^\dagger$, Addison W. Bohannon$^\star$}
\IEEEauthorblockA{$^\dagger~$UCLA, Los Angeles, CA 90095, USA,
Email: \{minedogan96, christina.fragouli\}@ucla.edu\\
$^*~$USC, Los Angeles, CA 90007, USA, Email: yessa@usc.edu\\
$^\star~$ARL, USARMY, Email: addison.w.bohannon.civ@mail.mil}

}

\maketitle

\begin{abstract}
We consider a source that wishes to communicate with a destination at a desired rate, {over a mmWave network where links are subject to blockage and nodes to failure (e.g., in a hostile military environment). To achieve resilience to link and node failures, we here explore} 
a state-of-the-art Soft Actor-Critic (SAC) deep reinforcement learning algorithm, that {adapts} the information flow through the network, without using knowledge of the link capacities or network topology. 
Numerical evaluations show that our algorithm can achieve the desired rate even in dynamic environments and it is robust against blockage. 
\end{abstract}

\section{Introduction}
Millimeter wave (mmWave) networks are expected to form a core part of 
5G and support a number of civilian and military applications; a number of use cases are currently built around {multi-hop} mmWave networks that range from private networks, such as in shopping centers, airports, museums and enterprises; mmWave mesh networks that use mmWave links as backhaul in dense urban scenaria; and military applications employing mobile hot spots,  remote sensing, control of unmanned aerial vehicles and surveillance of terrorist activities by full motion videos
\cite{Choi-Heath16,Mueck16,Brown18,mmWaveApps,Qualcomm19,Qualcomm-EU,Hur-13,Narita17,Woo-16}. 

But for the promise of these applications, it is well known that mmWave links are highly sensitive to blockage, channels may abruptly change and paths may get disrupted - and this is especially so for military applications~\cite{Jain,BaiTWCOM2015,MacCartneyGlobecom2017,WuTWCOM2021}.
Indeed,  in battlefields,  nodes can be destroyed, communication links can be highly volatile due to mobility of individual nodes, blockage can occur  
not only due to natural environment but also due to jamming, and it can be difficult and costly to replace nodes and repair damaged network connectivity~\cite{Burbank,Elmasry}.
Thus, we need  efficient
transmission mechanisms that can  fast adapt to blockages and abrupt channel variations, and offer throughput guarantees resilient to disruptions. 

In this paper, we explore the use of Deep Reinforcement Learning (DRL)  techniques, to gracefully adapt to blocked links and failed paths without collecting topology and channel knowledge. In particular, we consider a source that communicates with a destination over an arbitrary mmWave network, and ask to find which paths the source should use and at which rates to connect with the destination. A challenging aspect, captured through the 1-2-1 model that we introduced in~\cite{Ezzeldin_2018,EzzeldinISIT2019Multicast,EzzeldinISIT2019}, is that of scheduling:
in mmWave (and higher) frequencies,  due to high path loss, nodes communicate with each other by using beamforming, and steering their beams to connect to different neighbors: scheduling which nodes should communicate and for how long, is a non-trivial optimization problem.  In this paper, we resort to intelligent (RL) techniques, to gracefully adapt the schedule in the presence of disruptions, as a building stone towards autonomous network operation.

\noindent{\bf Related Work.} Several works in the literature propose relay selection schemes for mmWave networks \cite{dimas2019cooperative,8580768,7565000}, but focus on selecting the single path that has the highest signal-to-noise ratio (SNR), and thus do not offer resilience to blockage. 
A number of works, such as ~\cite{Kwon,He,Yuan,carlo} look at problems related to scheduling and mmWave beaforming, but require channel state information (CSI) and topology knowledge - our approach requires neither.  
Closer to ours is perhaps the work in~\cite{Jiang}, which uses multitask deep learning for multiuser hybrid beamforming-this work again relies on perfect knowledge of CSI and does not perform online learning. In this paper, we take advantage of deep learning, and in particular DRL algorithms, to select which paths to use in an online manner without using CSI or the network topology.

DRL algorithms use  deep neural networks as function approximators in large state and action spaces,
and have been widely employed in a multitude of applications such as games, robotics and communication networks \cite{Arulkumaran_2017, luong2018applications}. These algorithms  provide robust solutions in dynamic environments,   without requiring analytical models of large and complex systems \cite{luong2018applications}. A number of works  use RL for scheduling, routing and traffic engineering problems  \cite{wang2019deep,stampa2017deepreinforcement, xu2018experiencedriven}, but do not directly extend to mmWave networks.
Closer to ours are the works in~\cite{Xu} and~\cite{Vu}, that look at scheduling over mmWave networks using RL-based techniques - 
unlike ours, both approaches use CSI.


\noindent{\bf Contributions.} {
In our work, we leverage a state-of-the-art DRL algorithm called Soft Actor-Critic (SAC) algorithm \cite{haarnoja2018soft}, to support a desired  rate between a source and a destination over an arbitrary mmWave network.
To the best of our knowledge, this is the first time a DRL algorithm is employed for online optimization of multiple paths rates in mmWave networks. 
 Our DRL algorithm:\\
\noindent $\bullet$  Does not require knowledge of network topology or link capacities, and thus is well suited to volatile environments, such as encountered in military operations.\\
\noindent $\bullet$ Robustly adapts to link and node failures, as our evaluation results indicate, offering a superior performance to alternative algorithms we evaluate. }



\noindent \textbf{Paper Organization.} Section~\ref{sec:model} provides background on the 1-2-1 network model for mmWave networks and DRL. Section III explains the proposed algorithm. Section IV presents the evaluation results of the proposed algorithm. Section V concludes the paper.

\section{System Model and Background}\label{sec:model}
In this section, we provide background on Gaussian 1-2-1 networks and DRL.

\textit{Notation}.
$\abs{\cdot}$ is 
cardinality for sets, 
$\mathbb{E}\left[\cdot \right]$ denotes the expectation of a random variable, $\mathcal{H}\left(\cdot \right)$ is its entropy.
\subsection{Gaussian 1-2-1 Networks}
Gaussian 1-2-1 networks were introduced in~\cite{Ezzeldin_2018} to model information flow and study the capacity of multi-hop mmWave networks. In an $N$-relay Gaussian Full-Duplex (FD) 1-2-1 network, $N$ relays assist the communication between the source node (node $0$) and a destination node (node $N+1$).  Each FD node can simultaneously transmit and receive by using a single transmit beam and a single receive beam. 
Any two nodes have to direct their beams towards each other in order to activate a link that connects them.
Therefore, Gaussian 1-2-1 networks capture the steerable directivity of transmission in mmWave networks.

%
%
\noindent \textbf{Capacity of FD 1-2-1 networks.} In~\cite{Ezzeldin_2018}, the capacity of a Gaussian FD 1-2-1 network was approximated to within a constant gap that only depends on number of nodes in the network. In particular, it was shown in that the following Linear Program (LP) can compute the approximate capacity and the optimal beam schedule in polynomial-time:
\begin{align}
\label{capacity_paths}
\begin{array}{llll}
&\ \rm{P1:}\ \widebar{\msf{C}} = {\rm max}  \displaystyle\sum_{p \in \mathcal{P}} x_p \mathsf{C}_p   & & \\
&      ({\rm P1}a) \ x_p \geq 0 & \forall p \!\in\! \mathcal{P}, &  \\
&    ({\rm P1}b) \ \displaystyle\sum_{p \in \mathcal{P}_i}  x_p f^p_{p\pnext(i),i} \!\leq\! 1 & \forall i \! \in \! [0\!:\!N], & \\
&   ({\rm P1}c) \ \displaystyle\sum_{p \in \mathcal{P}_i} x_p f^p_{i,p\pprev(i)} \!\leq\! 1 & \forall i \! \in \! [1\!:\!N\!+\!1],  &
\end{array}
\end{align}
where: (i) $\mathcal{P}$ is the collection of all paths connecting the source to the destination; (ii) $C_p$ is the capacity of path $p$; (iii) $\mathcal{P}_i \subseteq \mathcal{P}$ is the set of paths that pass through node $i$ where $i \in [0:N+1]$; (iv) $p\pnext(i)$ (respectively, $p\pprev(i)$) is the node that follows (respectively, precedes) node $i$ in path $p$; (v) the variable $x_p$ is the fraction of time path $p$ is used; and (vi) $f^p_{j,i}$ is the optimal activation time for the link of capacity $\ell_{j,i}$ when path $p$ is operated, i.e.,
 $   f^p_{j,i} = C_p/\ell_{j,i}.$
Here, $\ell_{j,i}$ denotes the capacity of the link going from node $i$ to node $j$ where $(i,j) \in  [0:N]\times[1:N+1]$.
We refer readers to~\cite{Ezzeldin_2018} for a more detailed description.
\begin{remark}
{\rm 
The aforementioned algorithm relies on the centralized knowledge of network link capacities  to find the optimal schedule.
In this paper, we discard the assumption of the centralized knowledge and consider an online approach for scheduling that relies on the interaction between a DRL agent and the network. 
}
\end{remark}

\subsection{Deep Reinforcement Learning}
In this section, we provide background on DRL and in particular on the state-of-the-art SAC  algorithm.

In RL, an agent observes an environment and interacts with it. At time step $t$, the agent at state $\mathbf{s}_t$ takes the action $\mathbf{a}_t$ and it moves to the next state $\mathbf{s}_{t+1}$ while receiving the reward $r(\mathbf{s}_t,\mathbf{a}_t)$. In RL settings, states represent the environment and the set of all states is called the state space $\mathcal{S}$. Actions are chosen from an action space $\mathcal{A}$, where we use $\mathcal{A}\left(\mathbf{s}_t\right)$ to denote the set of possible (valid) actions at state $\mathbf{s}_t$. Rewards are numerical values given to the agent according to its actions and the aim of the agent is to maximize the long-term cumulative reward \cite{10.5555/3312046}. At each time step, the agent follows a policy $\pi\left(\cdot \;\middle\vert\; \mathbf{s}_t\right)$ that is a distribution (potentially deterministic) over actions given the current state $\mathbf{s}_t$. For episodic RL, the agent interacts with the environment for a finite horizon $T$ to maximize its long term cumulative reward.

In a lot of interesting RL settings, the enormous - if not continuous - nature of state and action spaces renders the use of classical tabular methods prohibitively inefficient. Thus, function approximators and model-free RL techniques are  employed to deal with these shortcomings~\cite{Arulkumaran_2017}. Although such techniques can be successful on challenging tasks, they suffer from two major drawbacks: high sample complexity and sensitivity to hyperparameters. Off-policy learning algorithms are proposed to improve sample efficiency, but they tend to experience stability and convergence issues particularly in continuous state and action spaces. 

The state-of-the-art SAC algorithm was proposed in~\cite{haarnoja2018soft} to improve the exploration and solve these stability issues. Since large state-action spaces require function approximators, SAC  uses approximators for the policy and Q function. In particular, the algorithm uses five parameterized functional approximators: policy function $(\phi)$; soft Q functions $(\theta_1$ and $\theta_2)$; and target soft Q functions $(\bar{\theta}_1$ and $\bar{\theta}_2)$. The aim is to maximize the following objective function
\begin{align}
J(\pi) = \sum_{t = 0}^{T}\mathbb{E}\left[r\left(\mathbf{s}_t,\mathbf{a}_t\right)+\alpha\mathcal{H}\left(\pi\left(\cdot \;\middle\vert\; \mathbf{s}_t\right)\right)\right],
\label{sac_objective}
\end{align}
where: (i) $T$ is the horizon; and (ii) the temperature parameter $\alpha$ indicates the relative importance of the entropy term to the reward. The entropy term  enables SAC algorithm to achieve improved exploration, stability and robustness~\cite{haarnoja2018soft}. 


\section{Proposed Reinforcement Learning Method}
In this section, we explain our network structure, RL formulation and the proposed algorithm.
\subsection{Network structure and RL environment}

We consider a Gaussian 1-2-1 network with an arbitrary topology where $N$ relay (intermediate) nodes operating in FD mode assist the communication between a source node (node 0) and a destination node (node $N+1$).
We assume that the channel coefficients (and as a consequence the link capacities) are unknown and they can change over time.
Our aim is to reach a certain desired rate $R^\star$ by using a small subset of the possible paths $\mathcal{P}_k$, where $\mathcal{P}_k \subseteq \mathcal{P}$ and $\left|\mathcal{P}_k\right| = k$. Therefore, our input and output size is equal to $k$.
The number $k$ affects the complexity of our algorithm. An implicit requirement is that, the selected set of paths $\mathcal{P}_k $ can jointly support the desired rate.  Selecting such a set of paths may require some network knowledge; yet we note that, unless we wish to operate a mmWave network close to its (very high) capacity, even randomly selected paths tend to meet this requirement.

The reason we aim to reach a  desired rate instead of  the full network capacity, is that, achieving the capacity may require a  large number of paths, which can increase the dimension of the problem substantially  and render the RL problem infeasible for large networks.
We formulate a single agent RL problem for this task by defining a Markov Decision Process (state space, action space and the reward function) as follows:

\noindent$\bullet$ \textit{State Space} $(\mathcal{S})$: Each state vector consists of rates of the selected paths. Therefore, if we denote the rate of the $i^{th}$ path in $\mathcal{P}_k$ at step $t$ by $x_{i,t}$, then the state vector at step $t$ is $\mathbf{s}_t = \left[x_{1,t}, x_{2,t},..., x_{k,t}\right] $.

\noindent$\bullet$ \textit{Action Space} $(\mathcal{A})$: Each action vector represents the changes in path rates of the selected paths. Formally, the action vector at step $t$ is $\mathbf{a}_t = \left[y_{1,t}, y_{2,t},..., y_{k,t}\right]$, where $y_{i,t}$ denotes the change in the path rate of the $i^{th}$ path in $\mathcal{P}_k$ at step $t$. Therefore, the state at step $t\!+\!1$ is 
    \[
    \mathbf{s}_{t+1} = \mathbf{s}_t+\mathbf{a}_t = \left[x_{1,t}\!+\!y_{1,t}, x_{2,t}\!+\!y_{2,t},..., x_{k,t}\!+\!y_{k,t}\right].
    \]

\noindent$\bullet$ \textit{Reward Function} $(r(\mathbf{s},\mathbf{a}))$: if the sum of the rates through the $k$ paths  exceeds the desired rate $R^\star$, the agent terminates the episode and receives a reward equal to $1$. Otherwise, the agent does not receive any rewards. 

The definitions of the state space, action space and reward function do not assume  knowledge of link capacities. We consider an episodic RL for this continuous control problem where each episode lasts for a finite horizon $T$. However, if the agent reaches the specified desired rate $R^\star$ during an episode, the episode ends early. Moreover, the next state has to be a physically feasible state, i.e., the fraction of time each path is used should satisfy the constraints of LP \rm{P1} in~\eqref{capacity_paths}. This means that all path rates have to be nonnegative, and a node cannot transmit or receive more than 100\% of the time. Therefore, if the action vector makes the next state invalid, it is assumed that the agent stays at the current state. We assume that the environment can determine whether a state is valid or not since if the path rates at a particular state are invalid, the network cannot support those rates and enters outage (for instance, drops packets due to queue congestion).
In a real deployment, the source node can be the agent, and it can observe the rates of the selected paths through TCP feedback and by observing packet drops. It can accordingly adjust the path rates at each step, and if it reaches the desired rate, terminate the episode.

In summary, the agent updates the path rates of the selected paths at each step through the action vector unless the action vector makes the next state invalid. In the latter case, the path rates are not changed and the agent stays at the current state. The proposed algorithm is provided in Algorithm \ref{scheduling_alg}. 
\begin{algorithm}
 \caption{Proposed Scheduling Algorithm}
 \label{scheduling_alg}
\begin{algorithmic}
\STATE {\bf Initialize:} SAC networks $\theta, \bar{\theta},\phi$
 as in \cite{haarnoja2018soft}.
\STATE {\bf Input:} Desired rate $R^\star$ and the set of paths $\mathcal{P}_k$.\\
\FOR{each episode}
    \STATE $\bullet$ Start with zero initial state, i.e., $\mathbf{s}_0 = \mathbf{0}$.
    \FOR{each environment step}
        \STATE $\bullet$ Get action $\mathbf{a}_t$ from policy $\pi_{\phi}(\mathbf{a}_t\mid \mathbf{s}_t)$. 
        \STATE $\bullet$ If the action makes the next state valid, move to the state $\mathbf{s}_{t+1} = \mathbf{a}_t+\mathbf{s}_{t}$. Otherwise, stay at the current state, i.e., $\mathbf{s}_{t+1} = \mathbf{s}_t$.
        
        \IF{the current rate is greater than or equal to $R^\star$ }
            \STATE $\bullet$ Receive reward $r_t = 1$.
            \STATE $\bullet$ Store the tuple $\left(\mathbf{s}_t,\mathbf{a}_t,r_t,\mathbf{s}_{t+1}\right)$ in the replay buffer.
            \STATE $\bullet$ Terminate the episode.
        \ELSE
            \STATE $\bullet$ Receive reward $r_t$ = 0.
            \STATE $\bullet$ Store the tuple $\left(\mathbf{s}_t,\mathbf{a}_t,r_t,\mathbf{s}_{t+1}\right)$ in the replay buffer.
        \ENDIF%
    \ENDFOR
    \FOR{each gradient step}
    \STATE $\bullet$ Update the network parameters using gradient descent.
    \ENDFOR

\ENDFOR

\end{algorithmic}
\end{algorithm}

\section{Performance Evaluation}
Here, we numerically evaluate the proposed method with respect to different performance metrics, as we discuss next.
\subsection{Experiment Settings}
\noindent \textbf{Simulated Network.} We used the same network architecture and the hyperparameters in~\cite{haarnoja2018soft} for the SAC algorithm (Table~\ref{parameters} lists the hyperparameters). The source code of our implementation is available online\footnote{\nolinkurl{github.com/minedgan/SAC_scheduling}}.
\begin{table}
\begin{center}
\begin{tabular}{|p{4.5cm}||p{1.5cm}|}
\hline
  \textbf{Parameter}
  & \textbf{Value}
 \\
\hline
 \hline
 optimizer & Adam \\
 \hline
 learning rate & $3\cdot 10^{-4}$  \\
 \hline
 discount & $1$  \\
 \hline
 replay buffer size & $10^6$  \\
 \hline
 number of hidden layers (all networks) & 2\\
 \hline
 number of hidden units per layer & 256 \\
 \hline
 number of samples per minibatch & 32\\
 \hline
 nonlinearity & ReLU\\
 \hline
 target smoothing coefficient & 0.005\\
\hline
\end{tabular}
\end{center}
\caption{SAC hyperparameters
}
\label{parameters}
\end{table}

We considered a fully-connected Gaussian 1-2-1 network with $N = 15$ relay nodes. The capacity of each link is sampled from a uniform distribution between $0$ and $10$.  
We considered both a static  and a time-varying network case. In the static case, the link capacities stayed constant throughout training, and the approximate capacity of the generated network was equal to $9.48$ as computed using the approach in~\cite{Ezzeldin_2018}. In the time-varying case, the capacity of each link was changed by a small random amount after each episode. This amount was sampled from a uniform distribution between $-1$ and $1$ (the average link capacity was $5.12$). The variation in the link capacities captures mobility of nodes or varying channel conditions. Among all possible paths\footnote{There were in total $35546\times10^{8}$ possible paths in the network.}, we used  $15$ paths to reach a desired rate - the remaining paths were not used. In the static case, the desired rate $R^\star$ was set to $60\%$ of the network capacity and  stayed constant during training. In the time-varying case, the desired rate $R^\star$ was set to $50\%$ of the network capacity. Since the link capacities changed in every episode, the network capacity and the desired rate $R^\star$ also changed accordingly. We  note that although in our experiments we calculated the capacity $\widebar{C}$ to display the performance, this is not  needed in a real deployment: as our evaluation shows, if a desired rate is achievable, a good agent  achieves it.

We are particularly interested in evaluating whether our algorithm can offer robust solutions in the presence of blockage. To create a realistic scenario, we considered a different blockage probability for each link, to capture  effects such as the link length. 
In particular, we assigned a randomly generated weight  $w_{j,i}$ to the link connecting node $i$ to node $j$   to represent its length, where $(i,j) \in  [0:N]\times[1:N+1]$. The weights were generated between $0$ and $250$  and they were symmetric, i.e., $w_{j,i} = w_{i,j}$, $\forall (i,j) \in  [1:N]\times[1:N]$.
{We also assumed that the arrival process of blockers is Poisson as in~\cite{Jain} and thus, we blocked the links between node $i$ and node $j$ with probability $1-e^{-\lambda w_{i,j}}$ where $\lambda$ subsumes the effects of blocker density and velocity of blockers. We chose $\lambda = 1/500$ in our experiments so that a high fraction of links could be blocked.} Note that if the weight (length) of a link is higher, the probability of that link being blocked is higher as well. During training, at every $10$ episodes, a new set of links was blocked and these links remained blocked until a new set was selected. For the blockage evaluations, we again considered the time-varying network, and the desired rate was set to $50\%$ of the network capacity in every episode.

The dimension of the state and action spaces is equal to the number of selected paths $k = 15$.  Among the $15$ paths, we included both high and low capacity paths to ensure that the desired rate could be supported and make our evaluation more realistic\footnote{There were $4$ high capacity paths among the selected $15$ paths.}. 
At a specific time, the network rate is calculated as the sum of rates of these selected $15$ paths.

We trained the agent for $200$ episodes and each episode had $T = 500$ time horizon, i.e., each episode lasted at most $500$ time steps. At each time step, an action vector was sampled from the policy which had a Gaussian distribution in our experiments. After sampling the action vector, we applied the hyperbolic tangent function (tanh) to the sample in order to bound the actions to a finite interval~\cite{haarnoja2018soft}. Moreover, we performed action clipping such that if the elements of the action vector (after applying tanh function) were less than $10^{-3}$, these elements were taken as $0$'s. The action clipping is necessary because the agent might need to take zero actions for some paths, particularly if these paths are blocked, and it is not possible to instantiate zero action by  sampling from a continuous distribution without action clipping.

\noindent \textbf{Baseline Methods.} Although there exists a
considerable number of routing algorithms in the literature, we cannot compare our proposed method with them: these existing algorithms are tailored to general networks and do not consider the scheduling constraints on mmWave networks, or they rely on knowledge of link capacities. Therefore, we compared our proposed method with the following two baselines.\\
{
$\bullet$  \textbf{Shortest Paths (SP)}: We found the shortest two paths out of $k=15$ paths by using the lengths $w_{j,i}$'s. Since the blockage probability of a link is high if the length of the link is high, we expect the probability of two shortest paths being blocked to be small. We should note that we did not select a single path whose length was the smallest since the network would not be able to reach the desired rate if this single path was blocked. Due to the scheduling constraints on mmWave networks, we performed equal time sharing across these two paths. \\
$\bullet$  \textbf{Equal Time Sharing (ES)}: We exploited all $15$ paths by performing equal time sharing across them due to the scheduling constraints. 
}
\begin{figure*}
     \centering
     \begin{subfigure}[b]{0.32\textwidth}
         \centering
         \includegraphics[width=\textwidth]{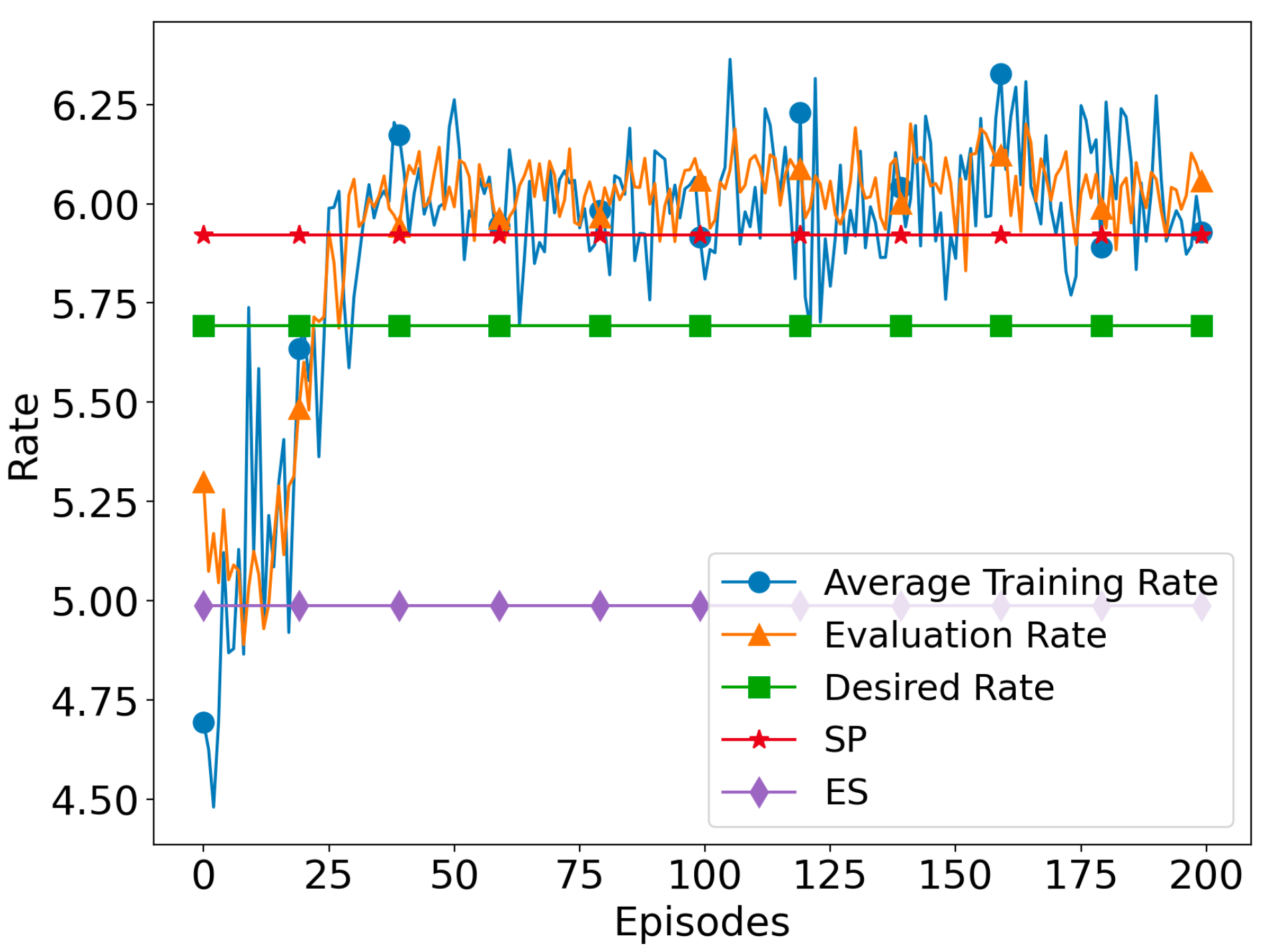}
         \caption{Static network.}
         \label{const}
     \end{subfigure}
     \begin{subfigure}[b]{0.32\textwidth}
         \centering
         \includegraphics[width=\textwidth]{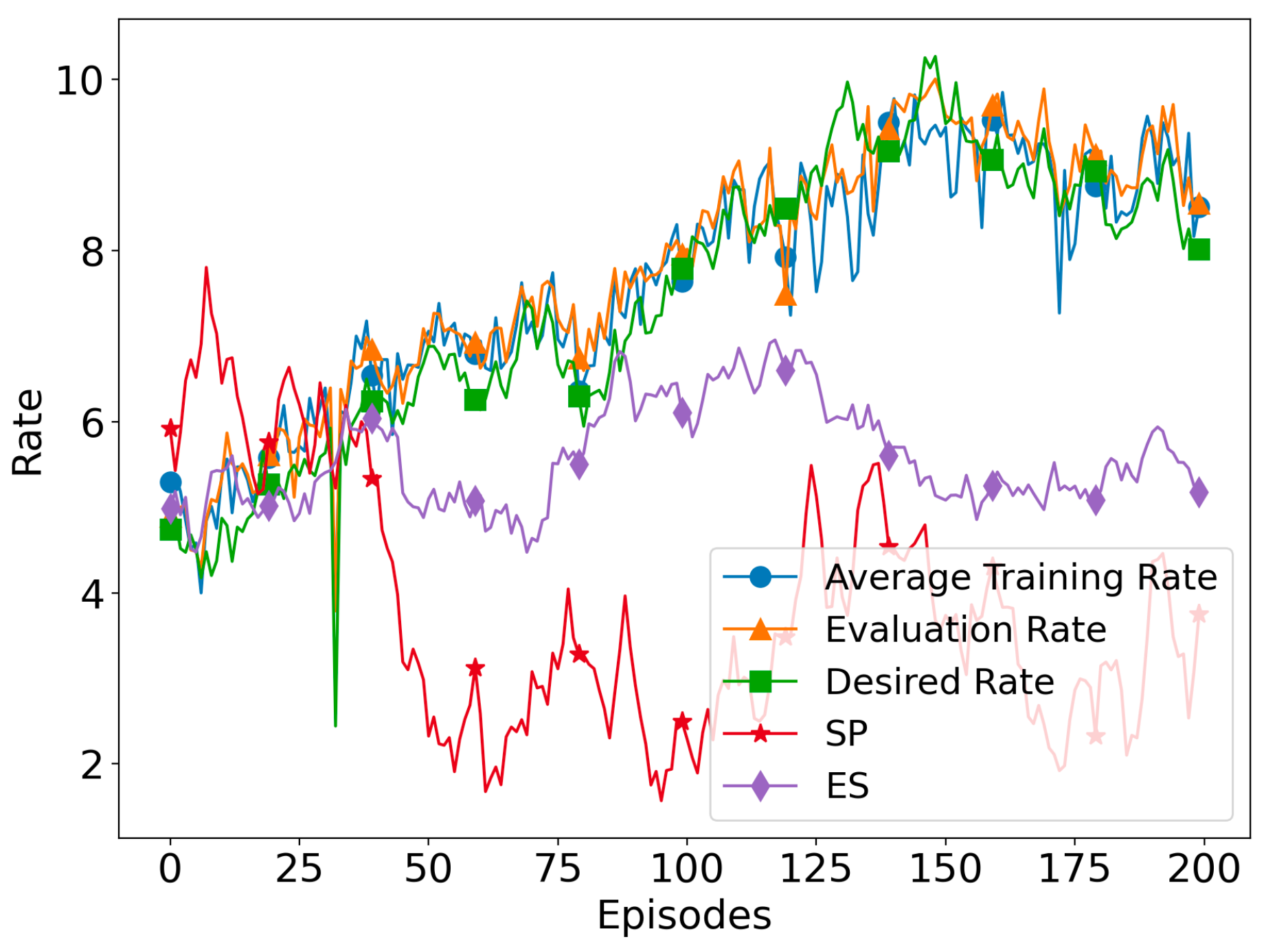}
         \caption{Time-varying network.}
         \label{timevarying}
     \end{subfigure}
     \begin{subfigure}[b]{0.32\textwidth}
         \centering
         \includegraphics[width=\textwidth]{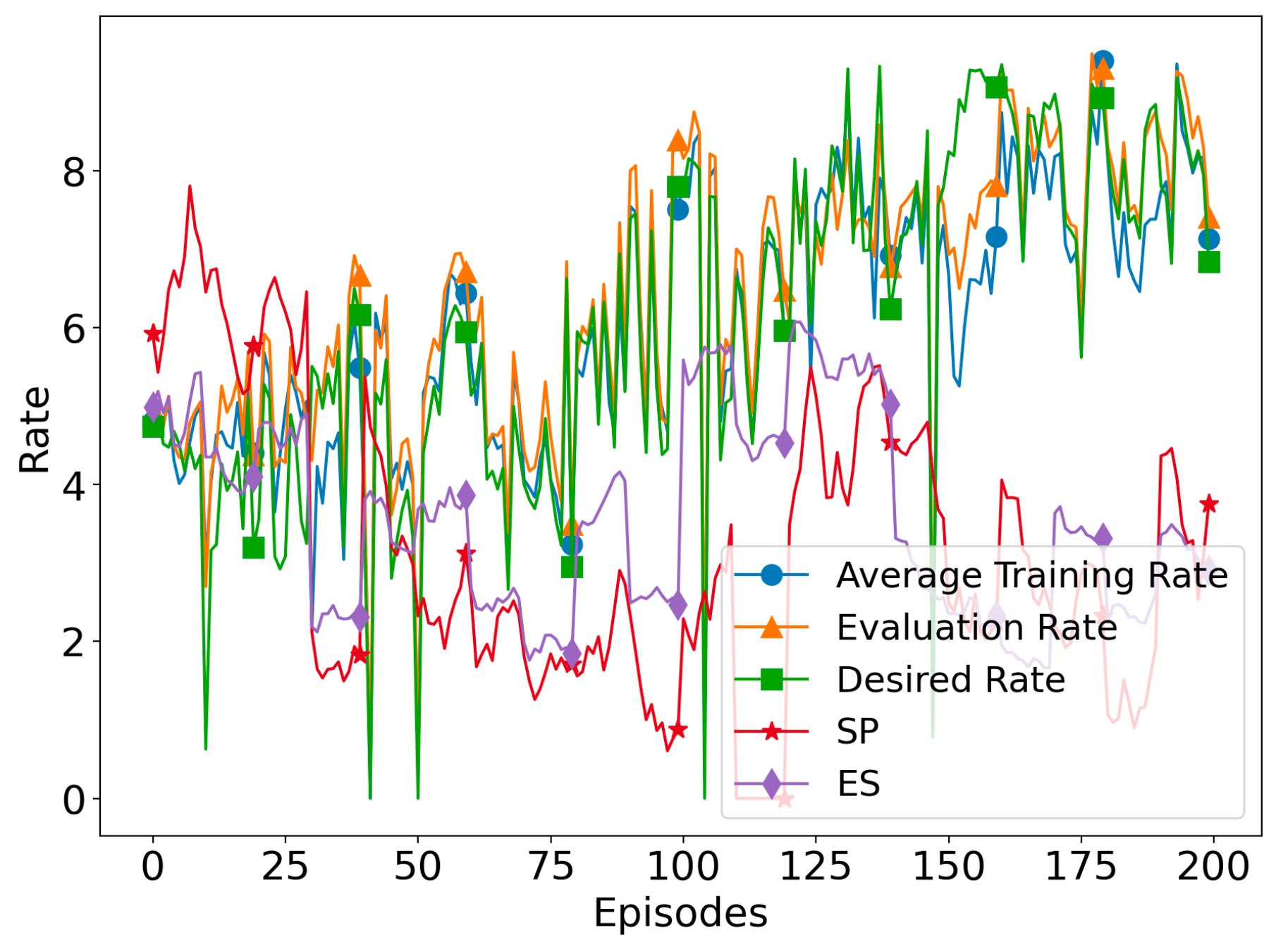}
         \caption{Blockage in the time-varying network.}
         \label{blockage}
     \end{subfigure}
        \caption{Performance of the proposed algorithm and baseline algorithms.}
        \label{comparison}
        \vspace{-0.1in}
\end{figure*}

\noindent \textbf{Performance Metrics.}
We evaluated the performance of the proposed algorithm by using the following two metrics.

\noindent $\bullet\ ${\em Average Training Rate.}  This is the average  rate achieved during training, which has important practical implications:  it captures whether and how fast the network  is able to support reasonable rates while training; it thus indicates whether it is possible to perform training online, while still utilizing the network. Towards this end, we trained five different instances of the algorithm with different random seeds and for each instance, we examined the average rate achieved in every episode during training. We found the average rate achieved in an episode by taking the average of the rates achieved at each time step during that episode. For episodes terminating earlier than the time horizon, the last rate is assumed to be maintained for the rest of the episode.
We finally took the average of the rates over these five instances. 

\noindent $\bullet\ ${\em Evaluation Rate.} At the end of each training episode, we performed evaluation and found the rate achieved by the agent.
Thus, this process could be considered as the validation of the policy during training. While finding the evaluation rate at the end of an episode, the agent started from zero initial state and adjusted the path rates by using its current policy. We again used $T=500$ time horizon: if the agent exceeded the desired rate during the evaluation, it stopped; otherwise, the final rate was the rate achieved at the last time step $t=500$. We should note that the policy of the agent in SAC algorithm is stochastic, therefore the agent can take different actions even at the same state. Thus, we repeated the same evaluation procedure for five times and took the average of the final rates to find the evaluation rate at that episode.

\subsection{Evaluation}
We here compare through simulation results the performance of our proposed algorithm
 versus the baseline algorithms SP and ES.
 Fig.~\ref{comparison} plots the achieved rates for the static case, the time-varying case, and the time-varying case with blockage. 
We observe the following.



\noindent $\bullet$ {\em Static network. } 
As shown in Fig.~\ref{const}, ES could not reach the desired rate. Indeed, ES performs equal time-sharing across all 15 paths which is not an optimal schedule: to support high rates, we need to activate the high capacity paths for a longer time and the low capacity paths for a shorter time, while still satisfying the constraints of LP P1 in \eqref{capacity_paths}. This illustrates that a naive (suboptimal) schedule can result in a significant waste of resources. On the other hand, SP reached the desired rate despite the equal time sharing approach since the two shortest paths were strong enough to support the desired rate. 
Finally, the proposed algorithm exploited all $15$ paths and adjusted their rates following a  schedule that allowed to reach the desired rate.

\noindent$\bullet$ {\em Time-varying network.}
In Fig.~\ref{timevarying}, we considered the time-varying network case (without blockage). Neither ES nor SP supported the desired rate. As discussed earlier, ES uses equal time sharing across $15$ paths, thus wasting  resources on low capacity paths.
SP also fails in this case, because over a volatile network the $2$ paths used may not remain strong enough to support the desired rate. 
On the other hand, our proposed algorithm adjusts the path rates at every time step to find a good schedule and exploits all paths as necessary. Thus, it reached the desired rate despite the channel variations.

\noindent$\bullet$ {\em Blockage.}
In Fig.~\ref{blockage}, we considered the time-varying network with blockage. 
Fig.~\ref{blocked_paths} shows that a good fraction of the paths (up to $10$ out of the selected $15$)  was blocked in every episode.
\begin{figure}
	\centering
     \includegraphics[width=0.67\columnwidth]{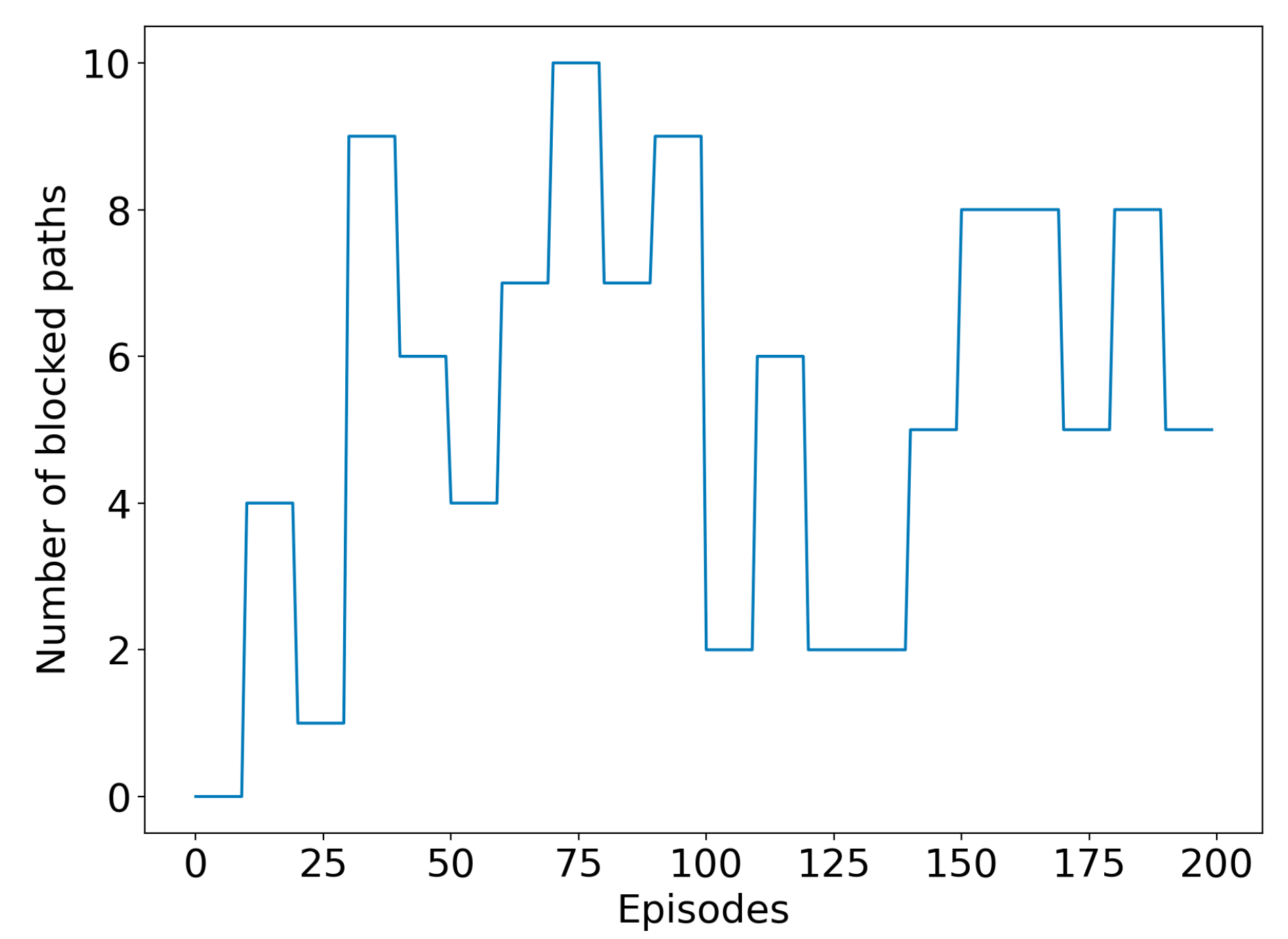}
     \caption{The number of blocked paths out of $15$ paths in Fig.~\ref{blockage}.}
     \label{blocked_paths}
     \vspace{-0.2in}
\end{figure}
As shown in Fig.~\ref{blockage}, neither ES nor SP supported the desired rate.
Indeed, neither ES nor SP adapt to blockage - if a path they use gets blocked, the corresponding time slot remains idle which results in wasted resources.  We note that SP uses the two paths with the smallest probability of blockage (shortest paths); yet blockage may still occur in one of them, which renders the target rate unattainable. 
On the other hand, our proposed approach adjusts the path rates such that it adapts to blockage and effectively exploits the unblocked paths. We should note that the agent does not use any side information - understanding if there are blockages in the network and bypassing the blocked paths is a part of its learning process.

\noindent$\bullet$ {\em Different number of paths.}
In the previous experiments, the agent used $k=15$ paths. In Fig.~\ref{diff_paths}, we show the performance of our algorithm when we vary the value $k$ in the time-varying network without blockage. 
\begin{figure}
	\centering
     \includegraphics[width=0.67\columnwidth]{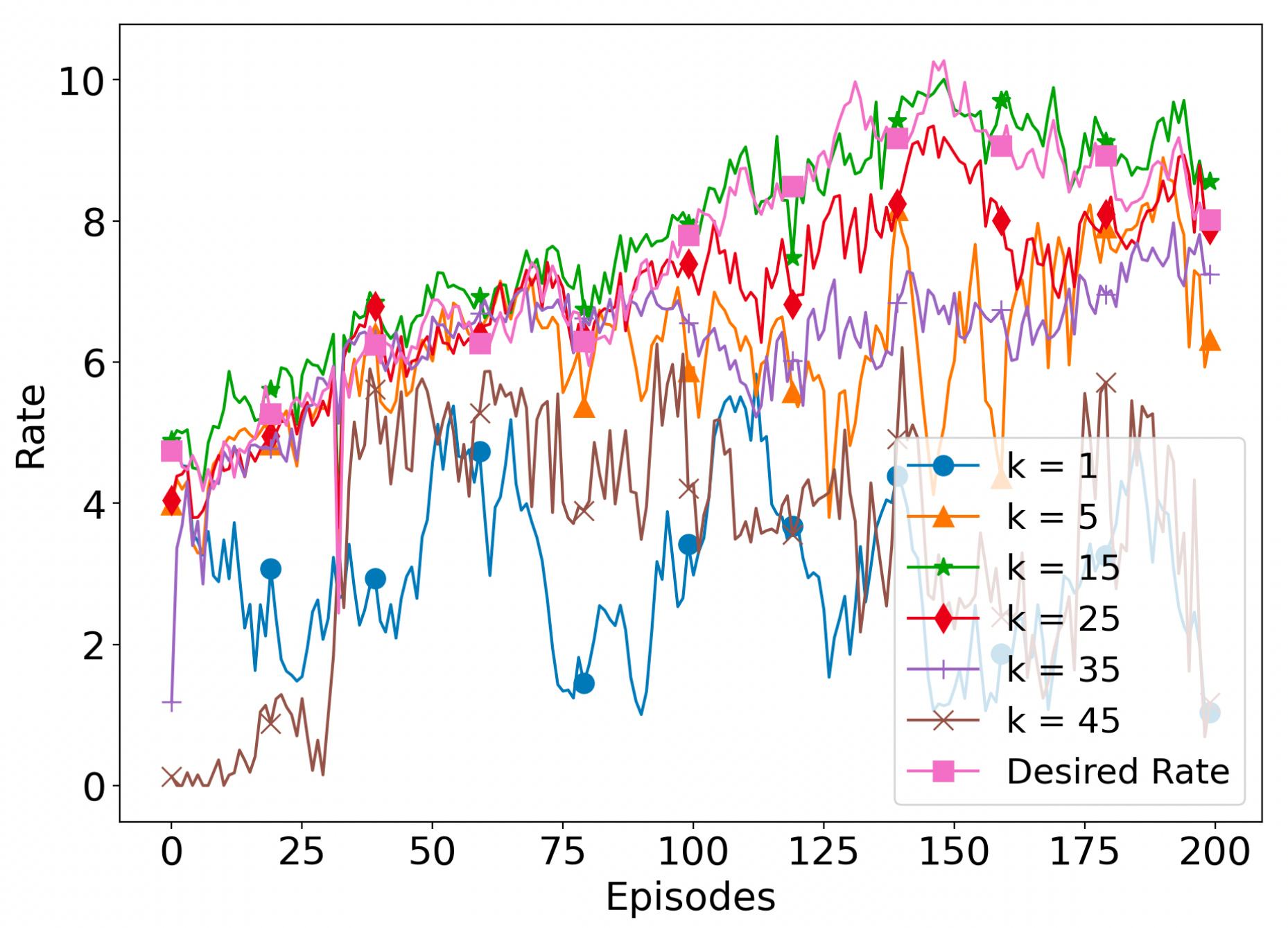}
     \caption{Performance of the algorithm for different values of $k$.}
     \label{diff_paths}
     \vspace{-0.2in}
\end{figure}
{We added new paths to the set $\mathcal{P}_k$ as we increased $k$.} The algorithm could not support the desired rate ($50\%$ of the capacity) by using $k=1$ or $k = 5$ paths since the selected paths were not strong enough to support the desired rate. Although the network supported the desired rate for $k=15$ paths, its performance degraded for higher values of $k$. As $k$ increases, the dimension of the continuous state and action spaces increases, and the agent needs to explore the space more effectively in order its function approximators to generalize well for unseen state-action pairs. 
{Thus, there is a trade-off: we may reach higher rates if a higher number of paths is used, however, this renders exploration more challenging.} Further exploring this trade-off is part of our future work. 

\section{Conclusions and Discussion}
In this paper, we started developing a DRL based approach to adaptively
select and route information over multiple paths in mmWave networks so as to achieve a desired source-destination rate. We formulated a single agent RL framework that does not assume any knowledge about the link capacities or the network topology. Our evaluations show that the proposed scheduling algorithm is robust against channel variations and blockage. Our work indicates that DRL techniques are promising and worth further exploration in the context of mmWave networks and our techniques form an encouraging first step towards robust and adaptable algorithms for military network scenaria, and a stepping stone towards autonomous network operation.



\bibliographystyle{IEEEtran}
\typeout{}
\bibliography{IEEEabrv,bibliography}

\begin{thebibliography}{10}
\providecommand{\url}[1]{#1}
\csname url@samestyle\endcsname
\providecommand{\newblock}{\relax}
\providecommand{\bibinfo}[2]{#2}
\providecommand{\BIBentrySTDinterwordspacing}{\spaceskip=0pt\relax}
\providecommand{\BIBentryALTinterwordstretchfactor}{4}
\providecommand{\BIBentryALTinterwordspacing}{\spaceskip=\fontdimen2\font plus
\BIBentryALTinterwordstretchfactor\fontdimen3\font minus
  \fontdimen4\font\relax}
\providecommand{\BIBforeignlanguage}[2]{{%
\expandafter\ifx\csname l@#1\endcsname\relax
\typeout{** WARNING: IEEEtran.bst: No hyphenation pattern has been}%
\typeout{** loaded for the language `#1'. Using the pattern for}%
\typeout{** the default language instead.}%
\else
\language=\csname l@#1\endcsname
\fi
#2}}
\providecommand{\BIBdecl}{\relax}
\BIBdecl

\bibitem{Choi-Heath16}
J.~{Choi}, V.~{Va}, N.~{Gonzalez-Prelcic}, R.~{Daniels}, C.~R. {Bhat}, and
  R.~W. {Heath}, ``Millimeter-wave vehicular communication to support massive
  automotive sensing,'' \emph{IEEE Communications Magazine}, vol.~54, no.~12,
  pp. 160--167, 2016.

\bibitem{Mueck16}
M.~{Mueck}, E.~C. {Strinati}, I.~{Kim}, A.~{Clemente}, J.~{Dore}, A.~{De
  Domenico}, T.~{Kim}, T.~{Choi}, H.~K. {Chung}, G.~{Destino}, A.~{Parssinen},
  A.~{Pouttu}, M.~{Latva-aho}, N.~{Chuberre}, M.~{Gineste}, B.~{Vautherin},
  M.~{Monnerat}, V.~{Frascolla}, M.~{Fresia}, W.~{Keusgen}, T.~{Haustein},
  A.~{Korvala}, M.~{Pettissalo}, and O.~{Liinamaa}, ``5{G} {CHAMPION} - rolling
  out 5{G} in 2018,'' in \emph{IEEE Globecom Workshops (GC Wkshps)}, 2016.

\bibitem{Brown18}
``What role will millimeter waves play in 5g wireless systems?''
  \url{https://www.mwrf.com/systems/what-role-will-millimeter-waves-play-5g-wireless-systems}.

\bibitem{mmWaveApps}
K.~Sakaguchi, T.~Haustein, S.~Barbarossa, E.~C. Strinati, A.~Clemente,
  G.~Destino, A.~P{\~{a}}rssinen, I.~Kim, H.~Chung, J.~Kim, W.~Keusgen, R.~J.
  Weiler, K.~Takinami, E.~Ceci, A.~Sadri, L.~Xian, A.~Maltsev, G.~K. Tran,
  H.~Ogawa, K.~M., and R.~W.~H. Jr., ``Where, when, and how mmwave is used in
  5{G} and beyond,'' \emph{IEICE Transactions on Electronics}, vol. E100.C,
  no.~10, pp. 790--808, 2017.

\bibitem{Qualcomm19}
``Qualcomm partners with {R}ussian mobile industry for mmwave 5{G} network in
  {M}oscow,'' \url{https://www.fiercewireless.com/5g/}.

\bibitem{Qualcomm-EU}
``Qualcomm introduces end-to-end over-the-air 5g mmwave test network in europe
  to drive 5g innovation,'' \url{https://www.qualcomm.com/news/releases/}.

\bibitem{Hur-13}
S.~{Hur}, T.~{Kim}, D.~J. {Love}, J.~V. {Krogmeier}, T.~A. {Thomas}, and
  A.~{Ghosh}, ``Millimeter wave beamforming for wireless backhaul and access in
  small cell networks,'' \emph{IEEE Transactions on Communications}, vol.~61,
  no.~10, pp. 4391--4403, 2013.

\bibitem{Narita17}
``Experience the next generation wireless {LAN} system, {W}i{G}ig contents
  download and viewing as {N}arita {A}irports new service trial!''
  \url{https://www.naa.jp/en/press/pdf/20170208-WiGig_en.pdf}.

\bibitem{Woo-16}
\BIBentryALTinterwordspacing
S.~Choi, H.~Chung, J.~Kim, J.~Ahn, and I.~Kim, ``Mobile hotspot network system
  for high-speed railway communications using millimeter waves,'' \emph{ETRI
  Journal}, vol.~38, no.~6, pp. 1052--1063, 2016. [Online]. Available:
  \url{https://onlinelibrary.wiley.com/doi/abs/10.4218/etrij.16.2716.0018}
\BIBentrySTDinterwordspacing

\bibitem{Jain}
I.~K. Jain, R.~Kumar, and S.~Panwar, ``Driven by capacity or blockage? a
  millimeter wave blockage analysis,'' in \emph{2018 30th International
  Teletraffic Congress (ITC 30)}, vol.~01, pp. 153--159.

\bibitem{BaiTWCOM2015}
T.~Bai and R.~W. Heath, ``Coverage and rate analysis for millimeter-wave
  cellular networks,'' \emph{IEEE Transactions on Wireless Communications},
  vol.~14, no.~2, pp. 1100--1114, 2015.

\bibitem{MacCartneyGlobecom2017}
\BIBentryALTinterwordspacing
G.~R. MacCartney, T.~S. Rappaport, and S.~Rangan, ``Rapid fading due to human
  blockage in pedestrian crowds at 5g millimeter-wave frequencies,'' in
  \emph{2017 {IEEE} Global Communications Conference}. [Online]. Available:
  \url{https://doi.org/10.1109/GLOCOM.2017.8254900}
\BIBentrySTDinterwordspacing

\bibitem{WuTWCOM2021}
Y.~Wu, J.~Kokkoniemi, C.~Han, and M.~Juntti, ``Interference and coverage
  analysis for terahertz networks with indoor blockage effects and
  line-of-sight access point association,'' \emph{IEEE Transactions on Wireless
  Communications}, vol.~20, no.~3, pp. 1472--1486, 2021.

\bibitem{Burbank}
J.~L. Burbank, P.~F. Chimento, B.~K. Haberman, and W.~T. Kasch, ``Key
  challenges of military tactical networking and the elusive promise of manet
  technology,'' \emph{IEEE Communications Magazine}, vol.~44, no.~11, 2006.

\bibitem{Elmasry}
G.~F. Elmasry, ``A comparative review of commercial vs. tactical wireless
  networks,'' \emph{IEEE Communications Magazine}, vol.~48, no.~10, 2010.

\bibitem{Ezzeldin_2018}
Y.~H. Ezzeldin, M.~Cardone, C.~Fragouli, and G.~Caire, ``Gaussian 1-2-1
  networks: Capacity results for mmwave communications,'' \emph{IEEE
  International Symposium on Information Theory (ISIT)}, 2018.

\bibitem{EzzeldinISIT2019Multicast}
Y.~H. {Ezzeldin}, M.~{Cardone}, C.~{Fragouli}, and G.~{Caire}, ``On the
  multicast capacity of full- duplex 1-2-1 networks,'' in \emph{IEEE
  International Symposium on Information Theory (ISIT)}, 2019.

\bibitem{EzzeldinISIT2019}
Y.~H. Ezzeldin, M.~Cardone, C.~Fragouli, and G.~Caire, ``Polynomial-time
  capacity calculation and scheduling for half-duplex 1-2-1 networks,'' in
  \emph{2019 IEEE International Symposium on Information Theory (ISIT)}, 2019,
  pp. 460--464.

\bibitem{dimas2019cooperative}
A.~Dimas, D.~S. Kalogerias, and A.~P. Petropulu, ``Cooperative beamforming with
  predictive relay selection for urban mmwave communications,'' \emph{{IEEE}
  Access}, 2019.

\bibitem{8580768}
Y.~{Yan}, Q.~{Hu}, and D.~M. {Blough}, ``Path selection with amplify and
  forward relays in mmwave backhaul networks,'' in \emph{IEEE PIMRC}, 2018.

\bibitem{7565000}
H.~{Abbas} and K.~{Hamdi}, ``Full duplex relay in millimeter wave backhaul
  links,'' in \emph{IEEE WCNC}, 2016.

\bibitem{Kwon}
G.~Kwon and H.~Park, ``A joint scheduling and millimeter wave hybrid
  beamforming system with partial side information,'' in \emph{2016 IEEE
  International Conference on Communications (ICC)}.

\bibitem{He}
S.~He, Y.~Wu, D.~W.~K. Ng, and Y.~Huang, ``Joint optimization of analog beam
  and user scheduling for millimeter wave communications,'' \emph{IEEE
  Communications Letters}, vol.~21, no.~12, 2017.

\bibitem{Yuan}
D.~Yuan, H.-Y. Lin, J.~Widmer, and M.~Hollick, ``Optimal joint routing and
  scheduling in millimeter-wave cellular networks,'' in \emph{IEEE INFOCOM 2018
  - IEEE Conference on Computer Communications}.

\bibitem{carlo}
H.~Shokri-Ghadikolaei, L.~Gkatzikis, and C.~Fischione, ``Beam-searching and
  transmission scheduling in millimeter wave communications,'' in \emph{2015
  IEEE International Conference on Communications (ICC)}.

\bibitem{Jiang}
J.~Jiang, Y.~Li, L.~Chen, J.~Du, and C.~Li, ``Multitask deep learning-based
  multiuser hybrid beamforming for mm-wave orthogonal frequency division
  multiple access systems,'' \emph{Science China Information Sciences},
  vol.~63, 08 2020.

\bibitem{Arulkumaran_2017}
K.~Arulkumaran, M.~P. Deisenroth, M.~Brundage, and A.~A. Bharath, ``Deep
  reinforcement learning: A brief survey,'' \emph{IEEE Signal Processing
  Magazine}, 2017.

\bibitem{luong2018applications}
N.~C. Luong, D.~T. Hoang, S.~Gong, D.~Niyato, P.~Wang, Y.~Liang, and D.~I. Kim,
  ``Applications of deep reinforcement learning in communications and
  networking: {A} survey,'' \emph{IEEE Communications Surveys \& Tutorials},
  2019.

\bibitem{wang2019deep}
J.~Wang, C.~Xu, Y.~Huangfu, R.~Li, Y.~Ge, and J.~Wang, ``Deep reinforcement
  learning for scheduling in cellular networks,'' in \emph{IEEE WCSP}, 2019.

\bibitem{stampa2017deepreinforcement}
G.~Stampa, M.~Arias, D.~S{\'a}nchez-Charles, V.~Munt{\'e}s-Mulero, and
  A.~Cabellos, ``A deep-reinforcement learning approach for software-defined
  networking routing optimization,'' \emph{arXiv preprint arXiv:1709.07080},
  2017.

\bibitem{xu2018experiencedriven}
Z.~Xu, J.~Tang, J.~Meng, W.~Zhang, Y.~Wang, C.~H. Liu, and D.~Yang,
  ``Experience-driven networking: {A} deep reinforcement learning based
  approach,'' in \emph{IEEE INFOCOM}, 2018.

\bibitem{Xu}
C.~Xu, S.~Liu, C.~Zhang, Y.~Huang, and L.~Yang, ``Joint user scheduling and
  beam selection in mmwave networks based on multi-agent reinforcement
  learning,'' in \emph{2020 IEEE 11th Sensor Array and Multichannel Signal
  Processing Workshop (SAM)}.

\bibitem{Vu}
T.~K. Vu, M.~Bennis, M.~Debbah, and M.~Latva-Aho, ``Joint path selection and
  rate allocation framework for 5g self-backhauled mm-wave networks,''
  \emph{IEEE Transactions on Wireless Communications}, vol.~18, no.~4, 2019.

\bibitem{haarnoja2018soft}
T.~Haarnoja, A.~Zhou, K.~Hartikainen, G.~Tucker, S.~Ha, J.~Tan, V.~Kumar,
  H.~Zhu, A.~Gupta, P.~Abbeel, and S.~Levine, ``Soft actor-critic algorithms
  and applications,'' \emph{ArXiv}, vol. abs/1812.05905, 2018.

\bibitem{10.5555/3312046}
R.~S. Sutton and A.~G. Barto, \emph{Reinforcement learning: An
  introduction}.\hskip 1em plus 0.5em minus 0.4em\relax MIT press, 2018.

\end{thebibliography}
\vspace{12pt}
\end{NoHyper}
\end{document}